\documentstyle[11pt,axodraw,a4]{article}

\newcommand\order{{\cal O}}
\renewcommand\Re{\mbox{Re}}
\renewcommand\Im{\mbox{Im}}
\newcommand\GeV{\mbox{ GeV}}
\newcommand\GEV{\mbox{GeV}}

\newcommand\PB{\mbox{pb}}

\begin{document}


\thispagestyle{empty}

\setcounter{page}{0}

\begin{flushright}
LMU-21/92\\
PSI-PR-93-05\\
TTP92-36\\
\end{flushright}
\vspace*{\fill}
\begin{center}
{\Large\bf $\order(\Gamma)$ Corrections to $W$ pair production\\
in $e^+e^-$ and $\gamma\gamma$ collisions}\\
\vspace{2em}
\large
\begin{tabular}[t]{c}
Andre Aeppli$^{a\dagger}$\\
Frank Cuypers$^b$\\
Geert~Jan van Oldenborgh$^c$\\
\\
{$^a$ \it Institut f\"ur Theoretische Teilchenphysik}\\
{\it Universit\"at Karlsruhe}\\
{\it Kaiserstr. 12, D-7500 Karlsruhe 1, Germany}\\
{$^b$ \it Sektion Physik der Universit\"at M\"unchen,}\\
{\it Theresienstra\ss e 37, D-8000 M\"unchen 2, Germany}\\
{$^c$ \it Paul Scherrer Institut, CH-5232 Villigen PSI, Switzerland}\\
\end{tabular}
\end{center}
\vspace*{\fill}

\begin{abstract}
Several schemes to introduce finite width effects to reactions involving
unstable elementary particles are given and the differences between them
are investigated.  The effects of the different schemes is investigated
numerically for $W$ pair production.  In $e^+e^-\to W^+W^-$ we find that
the effect of the non-resonant graphs cannot be neglected for
$\sqrt{s}\geq400\GeV$.  There is no difference between the various
schemes to add these to the resonant graphs away from threshold,
although some violate gauge invariance.  On the other hand, in the
reaction $\gamma\gamma\to W^+W^-$ the effect of the non-resonant graphs
is large everywhere, due to the $t$-channel pole.  However, even
requiring that the outgoing lepton is observable ($p_\perp >
.02\sqrt{s}$) reduces the contribution to about 1\%.  Again, the scheme
dependence is negligible here.
\end{abstract}
\vspace*{\fill}

\noindent$^\dagger${\footnotesize Supported by BMFT Grant No. 055KAP94P1}
\newpage


\section{Introduction}

Although a general treatment of unstable particles in
quantum field theory has been
given a long time ago \cite{VeltmanUnstable},
the technical difficulties especially in the context of gauge invariance
are considerable and need careful investigations.
Recently, several authors
\cite{Stuart1,Sirlin,HVeltmanUnstable} proposed a gauge invariant
procedure to include consistently higher order corrections at the $Z$ pole.
As an example of a 2-resonance production process, finite width
effects up to one-loop have been calculated for $e^+e^- \to ZZ$
\cite{Denner&SackZZ}. Here we will present a
computation of the tree level cross section for $W$-pair
production in various schemes. This reaction introduces two new elements:
non-resonant diagrams diagrams can not be separated from
the resonance production diagrams in a gauge invariant way,
and (in contrast to the case of
$Z$ pair production) large unitarity cancellations are present
in $e^+e^-$ collisions, which may
dramatically enhance gauge violating terms in the cross section.

The differences between these various schemes constitute the
$\order(\Gamma)$ corrections to the reaction.  Note that these are
formally of the same order as the one-loop corrections ($\Gamma/M
\propto \alpha$), and these must thus be combined properly.  This will
be the subject of a forthcoming paper \cite{Andre&Geert&Daniel}.

We compare several methods used to compute the off-shell effects, some
of which are gauge variant or have unacceptable properties near
threshold.  Next we consider what the practical implications are for the
total cross section of $W$ pair production, both at LEP II and at
future linear $e^+e^-$ and $\gamma\gamma$ colliders.  (The latter are
constructed out of an $e^-e^-$ collider by Compton scattering of an intense
laser beam \cite{Ginsburg}.)
In many channels $W$ pair production is the major
source of events, either as a signal to study the properties of the $W$ boson
\cite{Belanger&Boudjema} or as background to more exotic processes
\cite{ggamma}.  In both cases a thorough understanding of the standard model
prediction is required.

The layout of this paper is as follows.  In section \ref{sec:method}
we list the various
schemes that can be used to evaluate cross sections for reactions
involving (charged) unstable elementary particles, and their
shortcomings.  Next we perform this
calculation for $e^+e^- \to W^+W^-$ (section \ref{sec:eeWW}) and
$\gamma\gamma\to W^+W^-$ (section \ref{sec:ggWW}), discussing the
applicability of the various approximations.


\section{Methods}
\label{sec:method}

There are various methods to compute the cross sections for reactions
involving unstable particles.  We will give a list of these, with
discussions of their merits, in this section.  For simplicity we first
discuss the schemes for a single unstable particle and give the
extension to two particles later.  The prototypical reaction will be
$e^+e^-\to W^+\mu^-\bar{\nu}_\mu$ with the $W^+$ considered stable.

The first approximation to a process whose leading contributions
factorizes into the production
and decay of unstable particles is the narrow width approximation.  One treats
the unstable particles as stable in a production cross section and multiplies
with the relevant branching ratios, obtaining
\begin{eqnarray}
\label{nwa}
    \sigma_{\rm NWA} & = & \frac{1}{2s} \int\! dP\!S_{e^+e^-\to W^+W^-}
        |{\cal M}_{e^+e^-\to W^+W^-}|^2
        \times \frac{\Gamma_{W^-\to\mu^-\nu_\mu}}{\Gamma_W}
\nonumber\\& = & \frac{1}{2s} \int\! dP\!S_{e^+e^-\to W^+W^-\to
	\mu^-\bar{\nu}_\mu}
        |{\cal M}_{e^+e^-\to W^+W^-}|^2
        \times \frac{|M_{W^-\to\mu^-\nu_\mu}|^2}{2m_W \Gamma_W}
\ ,
\end{eqnarray}
where $dP\!S_{ab\to cd}$ denotes the phase space element.

This fails to take into account terms of $\order(\Gamma)$; the corresponding
dimensionless parameter will usually be $\Gamma/m$ (about 1/40 for the $W$),
but can be $m^3\Gamma/\lambda(s,m_1^2,m_2^2)$ near a threshold for the
production of particles 1 and 2 in the expansion of the phase space factor
$\sqrt{\lambda}$.  The narrow width approximation is not defined
below the threshold for production of the unstable particle.

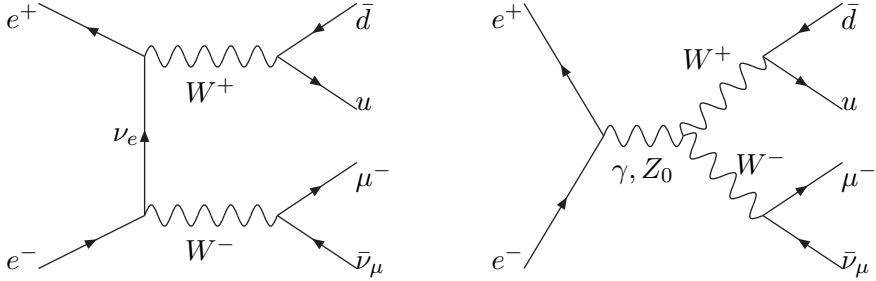
\begin{figure}[h]
\centerline{
\begin{picture}(180,100)(15,0)
\ArrowLine(0,0)(40,20)
\ArrowLine(40,20)(40,80)
\ArrowLine(40,80)(0,100)
\Photon(40,20)(90,20){4}{5}
\Photon(40,80)(90,80){4}{5}
\ArrowLine(120,0)(90,20)
\ArrowLine(90,20)(120,40)
\ArrowLine(120,100)(90,80)
\ArrowLine(90,80)(120,60)
\put(0,0){\makebox(0,0)[br]{$e^-$}}
\put(38,50){\makebox(0,0)[r]{$\nu_e$}}
\put(0,100){\makebox(0,0)[tr]{$e^+$}}
\put(65,12){\makebox(0,0)[t]{$W^-$}}
\put(65,72){\makebox(0,0)[t]{$W^+$}}
\put(120,0){\makebox(0,0)[bl]{$\bar{\nu}_\mu$}}
\put(120,40){\makebox(0,0)[tl]{$\mu^-$}}
\put(120,60){\makebox(0,0)[bl]{$u$}}
\put(120,100){\makebox(0,0)[tl]{$\bar{d}$}}
\end{picture}
\begin{picture}(180,100)(15,0)
\ArrowLine( 0, 0)(30,50)
\ArrowLine(30,50)(0,100)
\Photon(30,50)(60,50){4}{3}
\Photon(60,50)(90,20){4}{4}
\Photon(60,50)(90,80){4}{4}
\ArrowLine(120,0)(90,20)
\ArrowLine(90,20)(120,40)
\ArrowLine(120,100)(90,80)
\ArrowLine(90,80)(120,60)
\put(0,0){\makebox(0,0)[br]{$e^-$}}
\put(0,100){\makebox(0,0)[tr]{$e^+$}}
\put(45,42){\makebox(0,0)[t]{$\gamma,Z_0$}}
\put(79,80){\makebox(0,0)[r]{$W^+$}}
\put(80,36){\makebox(0,0)[bl]{$W^-$}}
\put(120,0){\makebox(0,0)[bl]{$\bar{\nu}_\mu$}}
\put(120,40){\makebox(0,0)[tl]{$\mu^-$}}
\put(120,60){\makebox(0,0)[bl]{$u$}}
\put(120,100){\makebox(0,0)[tl]{$\bar{d}$}}
\end{picture}
}
\caption[]{Resonant diagrams for $e^+e^-\to \mu^- \bar{\nu}_\mu u \bar{d}$}
\label{fig:eeWWdia_res}
\end{figure}

Above threshold the most important of these $\order(\Gamma)$ corrections
are included by simply using the on-shell expression for the matrix
element, but treating the kinematics and unstable particle propagator
off-shell.  This off-shell propagator is derived by resumming the
one-loop corrections to the propagator, giving rise to a simple
Breit-Wigner propagator $1/(p^2-m^2 + im\Gamma)$ with $\Gamma$ defined
by the relation\footnote{
  This approximation may not be sufficient in the threshold region and can
  be replaced by an s-dependent expression of the form
  $m\Gamma\rightarrow s\Gamma/m$.}
$m\Gamma = \Im\Pi(m^2)$ in terms of the self energy $\Pi$.  We refer to
this procedure as the ``resonant'' scheme:
\begin{eqnarray}
\label{eq:sigres}
    \sigma_{\rm res} & = & \frac{1}{2s} \int\! dP\!S_{e^+e^-\to
	W^+\mu^-\bar{nu}_\mu} \frac{ |{\cal M}_{e^+e^-\to W^+W^-}|^2
        \times |M_{W^-\to\mu^-\nu_\mu}|^2}{(p_{W^-}^2 - m_W^2)^2 +
	m_W^2 \Gamma_W^2}
\ .
\end{eqnarray}
The amplitude thus has the form
\begin{equation}
    {\cal M_{\rm res}} = \frac{R(m_W^2,\theta_i)}{p_{W^-}^2 - m_W^2 + i
	m_W \Gamma_W} \ ,
\end{equation}
where $R(p_{W^-}^2,\theta_i)/(p_{W^-}^2-m_W^2)$ denotes the amplitude
resulting from the resonant diagrams, i.e.\ those diagrams which contain
a $W$ which can be on-shell in the kinematically allowed region. (These
are given in Fig.\ \ref{fig:eeWWdia_res} for the full process.)
The essential variable in this amplitude
is the virtuality $p_{W^-}^2$ of the $W^-$.
The other kinematical variables are assumed to be angles, which are
independent of $p_{W^-}^2$.   We will suppress them from now on.
It can easily be verified that the narrow width
approximation follows from the limit $\Gamma\to0$ in Eq.\
(\ref{eq:sigres}).  Below threshold the resonant cross section is zero,
as $R(m_W^2) = 0$.

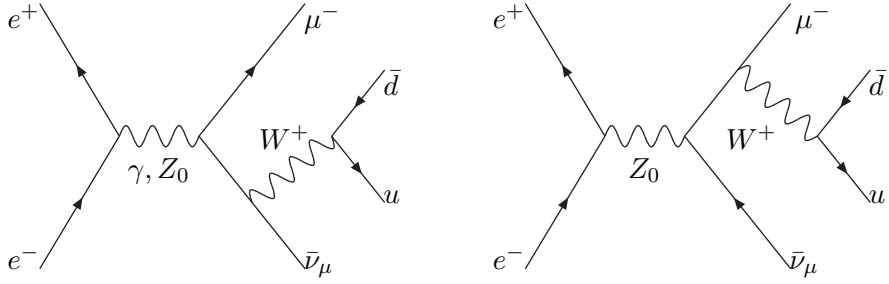
\begin{figure}
\centerline{
\begin{picture}(180,100)(-15,0)
\ArrowLine( 0, 0)(30,50)
\ArrowLine(30,50)(0,100)
\Photon(30,50)(60,50){4}{3}
\Line(100,0)(60,50)
\ArrowLine(60,50)(100,100)
\Photon(80,25)(110,50){4}{4}
\ArrowLine(130,75)(110,50)
\ArrowLine(110,50)(130,25)
\put(0,0){\makebox(0,0)[br]{$e^-$}}
\put(0,100){\makebox(0,0)[tr]{$e^+$}}
\put(45,42){\makebox(0,0)[t]{$\gamma,Z_0$}}
\put(100,0){\makebox(0,0)[bl]{$\bar{\nu}_\mu$}}
\put(100,100){\makebox(0,0)[tl]{$\mu^-$}}
\put(102,50){\makebox(0,0)[r]{$W^+$}}
\put(130,75){\makebox(0,0)[tl]{$\bar{d}$}}
\put(130,25){\makebox(0,0)[bl]{$u$}}
\end{picture}
\begin{picture}(180,100)(-15,0)
\ArrowLine( 0, 0)(30,50)
\ArrowLine(30,50)(0,100)
\Photon(30,50)(60,50){4}{3}
\ArrowLine(100,0)(60,50)
\Line(60,50)(100,100)
\Photon(80,75)(110,50){4}{4}
\ArrowLine(130,75)(110,50)
\ArrowLine(110,50)(130,25)
\put(0,0){\makebox(0,0)[br]{$e^-$}}
\put(0,100){\makebox(0,0)[tr]{$e^+$}}
\put(45,42){\makebox(0,0)[t]{$Z_0$}}
\put(100,0){\makebox(0,0)[bl]{$\bar{\nu}_\mu$}}
\put(102,100){\makebox(0,0)[tl]{$\mu^-$}}
\put(95,50){\makebox(0,0)[r]{$W^+$}}
\put(130,75){\makebox(0,0)[tl]{$\bar{d}$}}
\put(130,25){\makebox(0,0)[bl]{$u$}}
\end{picture}
}
\caption[]{Two of the four non-resonant diagrams for $e^+e^-\to \mu^-
\bar{\nu}_\mu u \bar{d}$}
\label{fig:eeWWdia_non}
\end{figure}

The next step is usually to also evaluate the matrix element off-shell,
thus to use an amplitude
\begin{equation}
    {\cal M_{\rm off}} = \frac{R(p_{W^-}^2)}{p_{W^-}^2 - m_W^2 + i m_W
	\Gamma_W}\ ;
\end{equation}
we refer to this procedure as the ``off-shell'' method.  This also gives
an answer below threshold.  In the case of
a charged resonance however, this procedure violates $U(1)_{em}$-gauge
invariance, even if the width is given by the physical (on-shell)
quantity, which is gauge invariant. The reason is that the original
resonant graphs  are not the only graphs which lead to the particular
final state:  non-resonant graphs (like those of
Fig.~\ref{fig:eeWWdia_non}) have to be included.  Only the sum of these
graphs is gauge invariant for $p_{W^-}^2 \neq m_W^2$ and an unresummed
unstable particle propagator.

The ``naive'' way to include these non-resonant graphs is to just add
them with a complex mass:
\begin{equation}
\label{eq:naive}
    {\cal M_{\rm naive}} = \frac{R(p_{W^-}^2)}{p_{W^-}^2 - m_W^2 + i m_W
	\Gamma_W} + N(p_{W^-}^2),
\end{equation}
where $N(p_{W^-}^2)$ denotes the contribution from the non-resonant
graphs.  However, as the amplitude was gauge invariant for $\Gamma_W=0$
it follows that it must be gauge variant for $\Gamma_W > 0$ (if the
resonant graphs are not separately gauge invariant).  We will discuss the
size of these gauge breaking terms at the end of this section.

An obviously gauge invariant procedure to include the non-resonant
graphs was introduced by Zeppenfeld {\em et al.}\ \cite{Zeppenfeld&Co}.
There, resonant and non-resonant terms  are
multiplied by the common factor $(p^2-m^2)/(p^2-m^2+im\Gamma)$:
\begin{equation}
\label{eq:Zep}
    {\cal M_{\rm all}} = \frac{R(p_{W^-}^2)}{p_{W^-}^2 - m_W^2 + i m_W
	\Gamma_W} + \frac{p^2-m^2}{p^2-m^2+im\Gamma} N(p_{W^-}^2),
\end{equation}
The price to save gauge invariance in this ``overall'' scheme is of
course an incorrect treatment of the non-resonant contribution close to
mass shell.   However, it can be argued that here the difference is of
higher order in this region of phase space.  We thus have a prescription
which is correct to leading order both on resonance and away from it,
but these contributions are formally of different order in $\alpha$.
Still, when the non-resonant graphs are large this may be a sensible
approximation.

Another gauge invariant way to include the off-shell effects is to
systematically separate orders in $\Gamma$.  A similar procedure was
used for instance in Ref.\ \cite{Stuart1} to include the higher-order
corrections at the $Z$-pole correctly.  In this ``polescheme'' the
matrix element has the form
\begin{equation}
\label{eq:pole}
    {\cal M_{\rm pole}} = \frac{R(m_W^2)}{p_{W^-}^2 - m_W^2 + i m_W
		\Gamma_W} + \Bigl[ \tilde{R}(p_{W^-}^2) + N(p_{W^-}^2) \Bigr]
\end{equation}
with
\begin{equation}
    \tilde{R}(p_{W^-}^2) = \frac{R(p_{W^-}^2) - R(m_W^2)}{p_{W^-}^2 - m_W^2}
\end{equation}
As the residue at the pole $p_{W^-}^2=m_W^2$ is gauge invariant one can
add the finite width in the first term without breaking gauge
invariance.  This corresponds to adding and resumming only the gauge
invariant part of the propagator corrections.  In this scheme
the cross section is given as the sum of the resonant cross section plus
$\order(\Gamma)$ corrections, which are both gauge invariant.  (The
half-resonant term $2\Re[R^\dagger(m_W^2)(\tilde{R}(p_{W^-}^2) +
N(p_{W^-}^2))/(p_{W^-}^2-m_W^2+im_W\Gamma_W)]$ and the non-resonant term
$|\tilde{R}(p_{W^-}^2) + N(p_{W^-}^2)|^2$ are of the same order in
$\Gamma_W$ after integration over $p_{W^-}^2$).  It is
thus a natural starting point for higher order corrections.  A detailed
discussion on the one-loop corrections will be given in a Ref.\
\cite{Andre&Geert&Daniel}.

However, the polescheme also has some undesirable properties.  The first
(resonant) term has a discontinuity when the threshold for the
production of the unstable particle is crossed.  Approaching from below,
one even encounters the original (non-resummed) singularity in the
propagator.  The accuracy of this scheme is thus doubtful around
threshold.

Below threshold we thus have the following unsatisfactory situation: the
narrow width scheme is undefined; the resonant cross section is zero and the
polescheme diverges.  The other schemes give a finite answer, but are
otherwise flawed.  It may be surmised that the problems in the threshold
region can be traced back to the low momenta of the $W$ bosons there, which
have time to exchange photons.  One does thus not expect a lowest order or
one-loop computation to give a reliable answer.  A bound-state calculation (as
recently performed for top production \cite{Kuhntop}) or resummation of the
resulting ladder graphs looks necessary here.

The  extensions for reactions with two unstable particles in the final
state are largely straightforward.  One now finds not only graphs of
$\order(\Gamma)$ but also of $\order(\Gamma^2)$; the doubly non-resonant
graphs.  In the case of $e^+e^-\to 4\rm\ fermions$,
these are only present
when there are electrons in the final state
In $\gamma\gamma$ collisions they are always present.
In the polescheme one can systematically neglect
these graphs; in the case of two similar particles like $W^+W^-$ the
$\order(\Gamma)$ corrections are just two times the size of the
corrections to the process with one particle regarded
unstable\footnote{If the decay products are treated identically
experimentally.}.

Below threshold one has to be very careful in the polescheme; in case of
the production of two identical particles (and
nothing else) can one regain a finite expression by symmetrizing:
\begin{eqnarray}
    \sigma & = & \int dP\!S_{e^+e^-\to4\mbox{\footnotesize\ fermions}} \frac{
        \theta(p_{W^-}^2 - p_{W^+}^2)}{p_{W^+}^2 - m_W^2}
        \Bigl| \frac{R(p_{W^+}^2,m_W^2)}{p_{W^-}^2 - m_W^2 + im_W\Gamma_W}
\nonumber\\&&\mbox{}
        + \frac{R(p_{W^+}^2,p_{W^-}^2) - R(p_{W^+}^2,m_W^2)}{p_{W^-}^2 -
        m_W^2} + N(p_{W^+}^2,p_{W^-}^2) \Bigr|^2 + ( W^+ \leftrightarrow W^-)
\end{eqnarray}
Of course this expression still diverges as $s\to4m_W^2$.

Finally, we compute the order of the difference between the three
schemes that include the non-resonant graphs.  The difference between
the ``overall'' scheme and the polescheme in the matrix element squared is
\begin{eqnarray}
    |{\cal M_{\rm pole}}|^2 - |{\cal M}_{\rm all}|^2 & = &
        -2 m_W \Gamma_W \quad \frac{ \Im\Bigl( R^\dagger(m_W^2)
            \bigl[ \tilde{R}(p_{W^-}^2) + N(p_{W^-}^2) \bigr] \Bigr)
        }{(p_{W^-}^2 - m_W^2)^2 + m_W^2 \Gamma_W^2}
\nonumber\\&&\mbox{}
        + m_W^2 \Gamma_W^2 \quad \frac{
          \Bigl| \tilde{R}(p_{W^-}^2) + N(p_{W^-}^2) \Bigr|^2
        }{(p_{W^-}^2 - m_W^2)^2 + m_W^2 \Gamma_W^2}
\end{eqnarray}
The first term, which is of $\order(\Gamma)$ relative to the resonant term, is
proportional to the Levi-Civita tensor $\epsilon_{\mu\nu\rho\sigma}p_1^\mu
p_2^\nu p_3^\rho p_4^\sigma$, with the $p_i^\mu$ four independent momenta.
This term will disappear when integrated over a set of reasonably symmetric
cuts\footnote{To be precise, the cuts should not introduce more than one
spatial direction other than the beam axis so that there are no more than
three independent four vectors after integration.}.  Note that this is not the
case in general in doubly differential cross sections.  The second term is of
$\order(\Gamma^2)$ and thus negligible in our approximations.

The difference between the gauge variant ``naive'' formulation and the
polescheme is given by
\begin{eqnarray}
    |{\cal M}_{\rm pole}|^2 - |{\cal M}_{\rm naive}|^2 & = &
        2 m_W \Gamma_W \quad \frac{
        \Im\Bigl( (p_{W^-}^2 - m_W^2) \tilde{R}^\dagger(p_{W^-}^2) N(p_{W^-}^2)
        - R^\dagger(m_W^2) \tilde{R}(p_{W^-}^2) \Bigr)
        }{(p_{W^-}^2 - m_W^2)^2 + m_W^2 \Gamma_W^2}
\nonumber\\&&\mbox{}
        + m_W^2 \Gamma_W^2 \quad \frac{
        \Bigl| \tilde{R}(p_{W^-}^2) \Bigr|^2
        + 2\Re\Bigl( \tilde{R}^\dagger(p_{W^-}^2) N(p_{W^-}^2) \Bigr)
        }{(p_{W^-}^2 - m_W^2)^2 + m_W^2 \Gamma_W^2}
\end{eqnarray}
Again, the $\order(\Gamma)$ term will disappear when integrated over a
symmetric part of phase space, and the difference (including the gauge
breaking terms) are of order $\Gamma^2$.

A summary of the six different schemes we have defined in this section
and their properties can be found in table \ref{table}.


\section{$e^+e^-\to W^+W^-$}
\label{sec:eeWW}

Depending on the cm energy $\sqrt{s}$ of this reaction, two complementary
aspects of the problem are revealed. Even at highest possible LEP II energies
\cite{ECFA_LEP200}
the inherent unitarity cancellations are at most of the order of a few percent
of the total cross section whereas the effect of the $W$-width amounts
to more than 10\%.  Testing the Standard model predictions within reasonable
limits at LEP II requires therefore an accuracy of the calculations below 1\%
and thus a detailed analysis of the width effects.  On the other hand at
energies $\sqrt{s} \approx 500$ GeV \cite{EE500} the finite width effect is
small but the unitarity cancellations account for more than an order of
magnitude. Any gauge breaking term induced by resummation or the omission of
non-resonant graphs may thus be enhanced and could upset any prediction of the
model.

In the following we will illustrate numerically the general
results obtained in the
previous section. For definiteness we consider
$e^+e^- \rightarrow u\bar{d}\mu\bar{\nu_{\mu}}$ in the $\alpha$-scheme;
therefore we use the fine structure constant $\alpha$, $M_{Z}=91.177$ GeV
and correspondingly $M_{W}=80.23$ GeV as input parameters.
The value for the $W$-width is then $\Gamma_{W}=2.072$ GeV,
assuming $m_{t}=140$ GeV and $m_{H}=100$ GeV \cite{DennerHabilitation}.
We work in the unitary gauge.

\begin{figure}
\begin{picture}(554,504)(50,0)
\put(0,0){\strut\epsffile{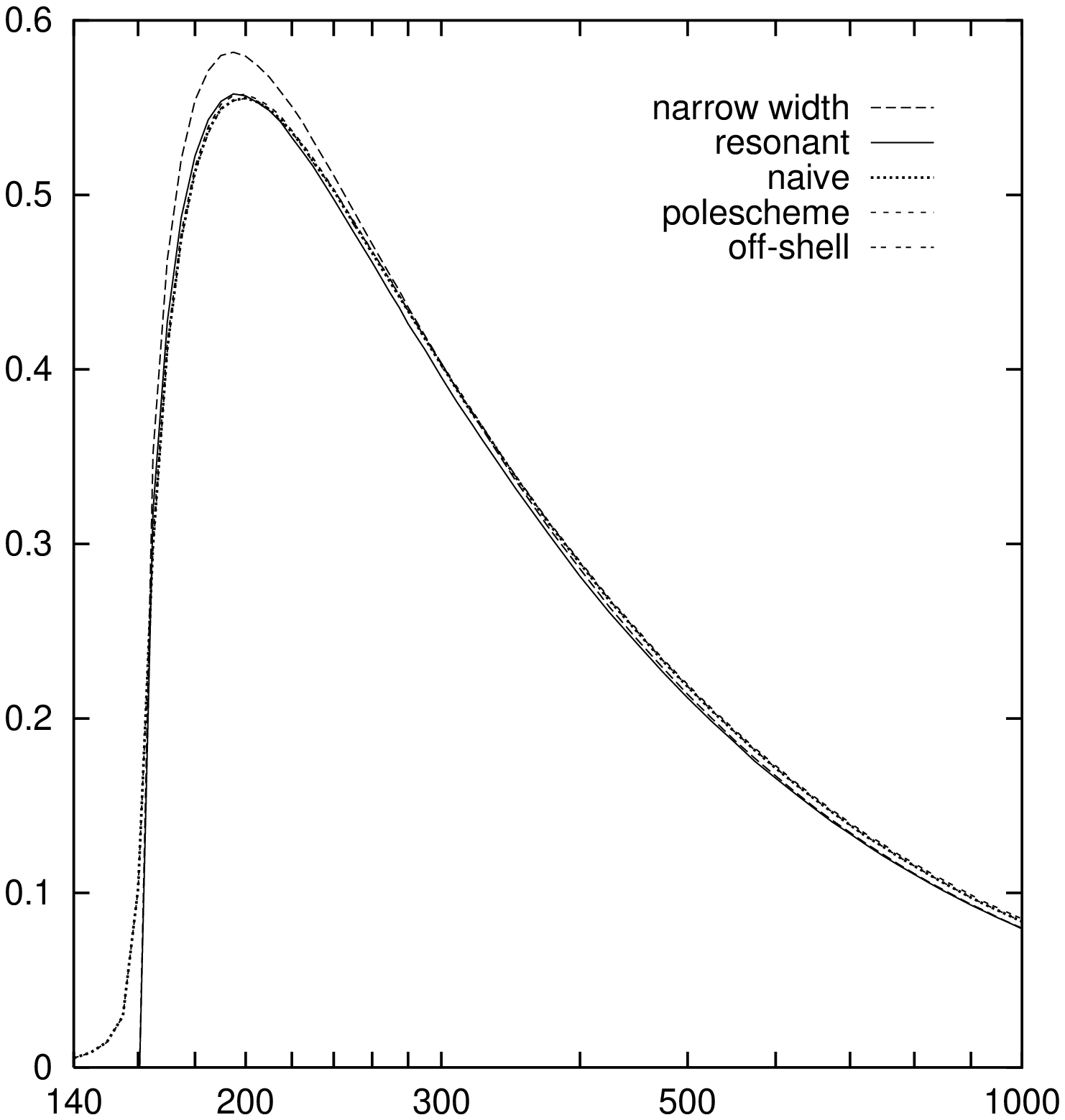}}
\put( 60.8,468.9){\makebox(0,0)[tr]{\shortstack{\Large\strut
$\sigma$\\\Large\strut$[\PB]$}}}
\put(480.8, 19.1){\makebox(0,0)[tr]{\Large$\sqrt{s_{ee}}[\GEV]$}}
\end{picture}
\caption[]{Cross sections for $e^+e^-\to \mu^- \bar{\nu}_\mu u \bar{d}$ in the
different schemes defined in the text.
The ``overall'' scheme yields results which are indistinguishable from
those of the ``naive'' scheme.}
\label{fig:eeWWabs}
\end{figure}

\begin{figure}
\begin{picture}(554,504)(50,0)
\put(0,0){\strut\epsffile{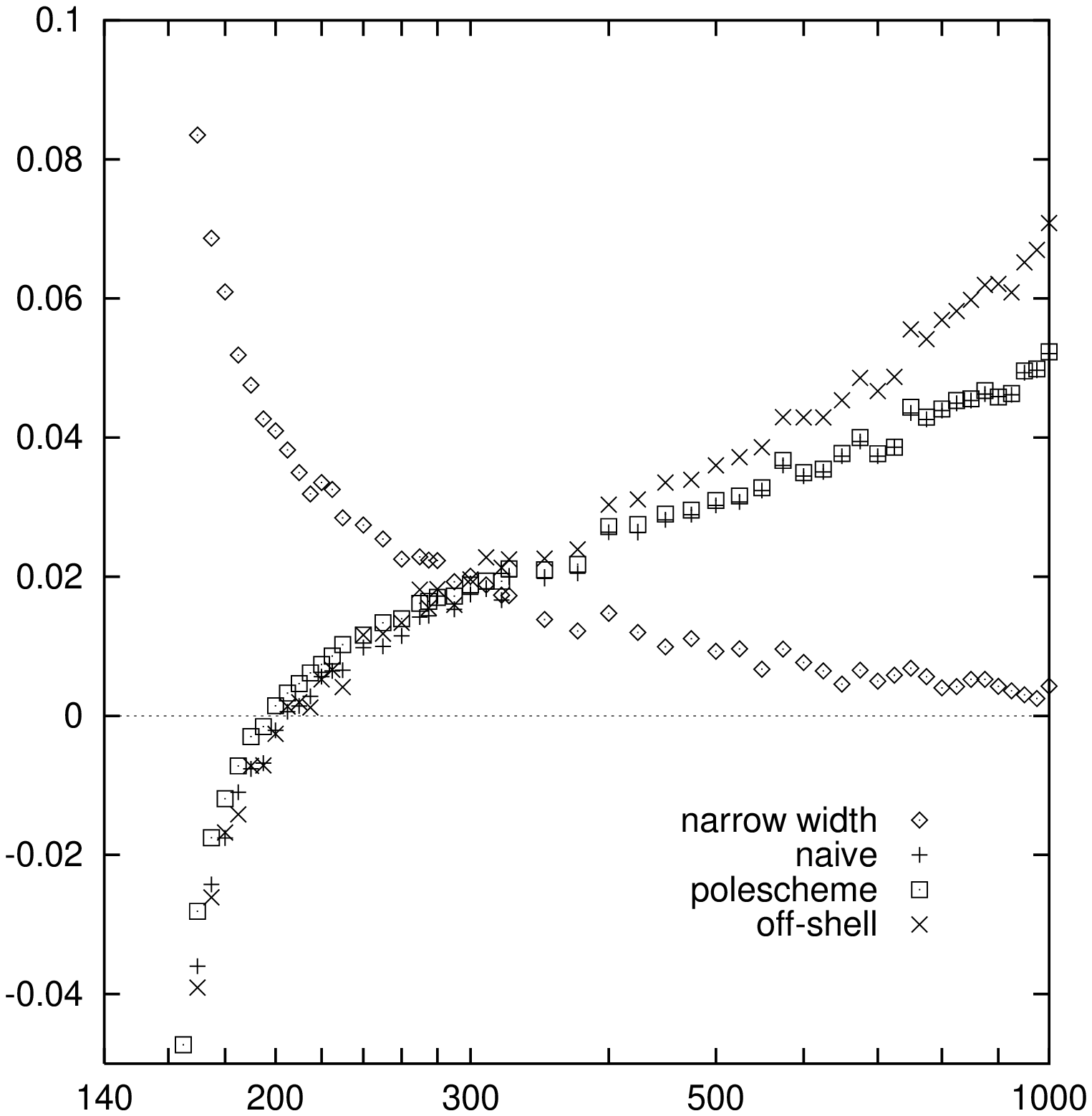}}
\put( 60.8,468.9){\makebox(0,0)[tr]{\Large\strut
$\frac{\displaystyle\sigma-\sigma_{\rm res}}{\displaystyle\sigma_{\rm res}}$}}
\put(480.8, 19.1){\makebox(0,0)[tr]{\Large$\sqrt{s_{ee}}[\GEV]$}}
\end{picture}
\caption[]{Relative differences with the doubly resonant cross section for
$e^+e^-\to \mu^- \bar{\nu}_\mu u \bar{d}$.  The scatter is due to limited
integration accuracy.}
\label{fig:eeWWfrac}
\end{figure}

The resonant diagrams are the well known s- and t-channel diagrams given in
Fig.\ \ref{fig:eeWWdia_res}. In addition there are many non-resonant diagrams
with essentially 3 different topologies. Imposing appropriate restrictions
on the 4 fermion final state (no electrons), only annihilation-type diagrams
(like in Fig.\ \ref{fig:eeWWdia_non}) contribute. Since the flavor content of
the final state does not alter gauge cancellation, it is clear that diagrams
with the other topologies form gauge invariant subsets by themselves.
The result for the resonant diagrams using a Breit-Wigner type propagator
has been given some time ago \cite{Muta}.
In Ref.\ \cite{Andre&DanielBG}, the annihilation type non-resonant diagrams
has been added.
In Fig.\ \ref{fig:eeWWabs} we show the results of an
explicit calculation of all the schemes discussed so far.
Except below threshold,
where only some schemes give a non-zero result,
and the narrow width approximation
near threshold, they all agree to within a few percent.  The differences with
the doubly resonant cross section are shown in Fig.\ \ref{fig:eeWWfrac}.

The difference between the gauge invariant ``overall'' scheme
and the ``naive'' results of \cite{Andre&DanielBG} is completely
negligible over this range of $\sqrt{s}$, at least in the unitary gauge
that we use.  The only sizable effect of the non-resonant diagrams is
obtained for energies close to threshold and above 400 GeV.
Otherwise,
taking the resonant diagrams off-shell and ignoring the non-resonant graphs
gives a very good description.

The two gauge invariant calculations
agree very well down to 180 GeV.
At $\sqrt{s}=185\GeV$ the difference is 0.4\%.
Closer to threshold
the two procedures start deviating significantly.
Since the ``overall'' scheme  result mistreats
the non-resonant terms,
whereas the polescheme is highly discontinuous,
we take this difference to indicate our ignorance of the threshold region.


\section{$\gamma\gamma\to W^+W^-$}
\label{sec:ggWW}

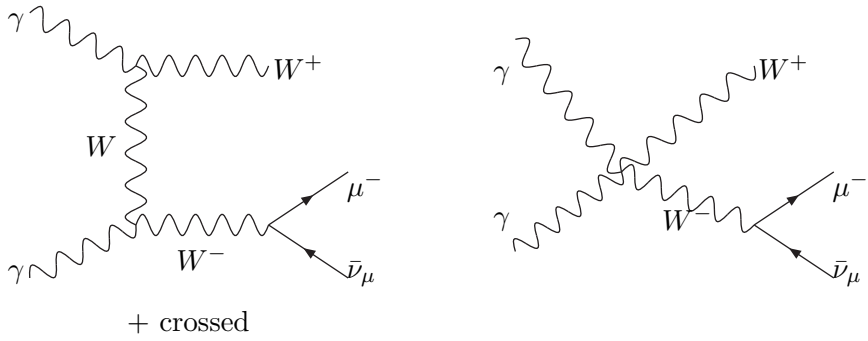
\begin{figure}
\centerline{
\begin{picture}(180,120)(-15,-20)
\Photon(0,0)(40,20){4}{4}
\Photon(40,20)(40,80){4}{5}
\Photon(40,80)(0,100){4}{4}
\Photon(40,20)(90,20){4}{5}
\Photon(40,80)(90,80){4}{5}
\ArrowLine(120,0)(90,20)
\ArrowLine(90,20)(120,40)
\put(-2,0){\makebox(0,0)[br]{$\gamma$}}
\put(33,50){\makebox(0,0)[r]{$W$}}
\put(-2,100){\makebox(0,0)[tr]{$\gamma$}}
\put(65,12){\makebox(0,0)[t]{$W^-$}}
\put(92,80){\makebox(0,0)[l]{$W^+$}}
\put(120,0){\makebox(0,0)[bl]{$\bar{\nu}_\mu$}}
\put(120,40){\makebox(0,0)[tl]{$\mu^-$}}
\put(60,-20){\makebox(0,0)[b]{$+$ crossed}}
\end{picture}
\begin{picture}(180,120)(-15,-20)
\Photon(0,10)(40,40){4}{5}
\Photon(0,90)(40,40){4}{5}
\Photon(90,20)(40,40){4}{5}
\Photon(90,80)(40,40){4}{5}
\ArrowLine(120,0)(90,20)
\ArrowLine(90,20)(120,40)
\put(-2,20){\makebox(0,0)[br]{$\gamma$}}
\put(-2,80){\makebox(0,0)[tr]{$\gamma$}}
\put(65,27){\makebox(0,0)[t]{$W^-$}}
\put(92,80){\makebox(0,0)[l]{$W^+$}}
\put(120,0){\makebox(0,0)[bl]{$\bar{\nu}_\mu$}}
\put(120,40){\makebox(0,0)[tl]{$\mu^-$}}
\end{picture}
}
\caption[]{Resonant diagrams for $\gamma\gamma\to \ell^- \bar{\nu}_\ell W^+$}
\label{fig:ggWWdia_res}
\end{figure}

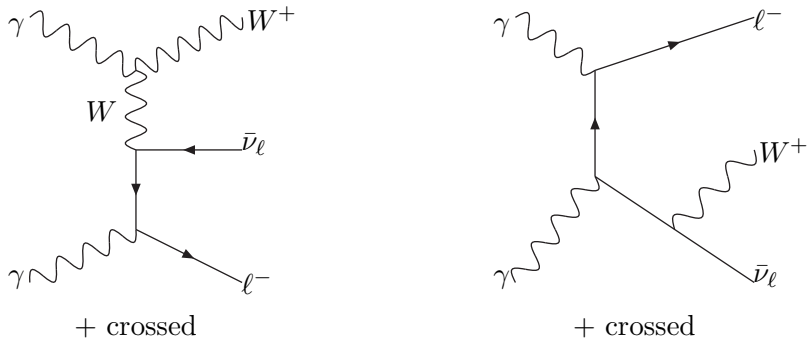
\begin{figure}
\centerline{
\begin{picture}(180,120)(-20,-20)
\Photon(0,0)(40,20){4}{4}
\Photon(40,50)(40,80){4}{3}
\Photon(40,80)(0,100){4}{4}
\Photon(40,80)(80,100){4}{5}
\ArrowLine(80,50)(40,50)
\ArrowLine(40,50)(40,20)
\ArrowLine(40,20)(80,0)
\put(-2,0){\makebox(0,0)[br]{$\gamma$}}
\put(33,65){\makebox(0,0)[r]{$W$}}
\put(-2,100){\makebox(0,0)[tr]{$\gamma$}}
\put(82,100){\makebox(0,0)[l]{$W^+$}}
\put(80,50){\makebox(0,0)[bl]{$\bar{\nu}_\ell$}}
\put(80,0){\makebox(0,0)[l]{$\ell^-$}}
\put(40,-20){\makebox(0,0)[b]{$+$ crossed}}
\end{picture}
\begin{picture}(180,120)(-20,-20)
\Photon(0,0)(30,40){4}{4}
\Photon(0,100)(30,80){4}{3}
\Line(90,0)(30,40)
\ArrowLine(30,40)(30,80)
\ArrowLine(30,80)(90,100)
\Photon(60,20)(90,50){4}{3}
\put(-2,0){\makebox(0,0)[br]{$\gamma$}}
\put(-2,100){\makebox(0,0)[tr]{$\gamma$}}
\put(92,50){\makebox(0,0)[l]{$W^+$}}
\put(90,0){\makebox(0,0)[bl]{$\bar{\nu}_\ell$}}
\put(90,100){\makebox(0,0)[l]{$\ell^-$}}
\put(45,-20){\makebox(0,0)[b]{$+$ crossed}}
\end{picture}
}
\caption[]{Non-resonant diagrams for $\gamma\gamma\to \ell^- \bar{\nu}_\ell
W^+$}
\label{fig:ggWWdia_non}
\end{figure}

The resonant and non-resonant Feynman diagrams for the
reaction $\gamma\gamma\to \ell^- \bar{\nu}_\ell W^+$ are given in
Figs \ref{fig:ggWWdia_res} and \ref{fig:ggWWdia_non} respectively.  To
restrict ourselves to the $\order(\Gamma)$ corrections we consider the
$W^+$ stable; otherwise doubly non-resonant diagrams are needed for a
gauge invariant result.  The total corrections with both $W$ bosons
unstable are about twice the size of the corrections given here.  (In
the polescheme the factor 2 is exact, up to $\order(\Gamma^2)$ terms).

This reaction is important at possible future $\gamma\gamma$ colliders
to study the $WW\gamma$ coupling and as a background to missing
$p_\perp$ physics \cite{ggamma}.  The non-resonant diagrams contain
large logarithms $\log(m_\ell^2/s) \Gamma_W/m_W$ in the total cross
section (as appear in the similar diagrams for the reaction
$\gamma\gamma\to e^+e^-Z_0$).  To suppress these we impose a $p_\perp$
cut on the outgoing leptons, $p_\perp(\ell) > 0.02 \sqrt{s}$; this
should give an indication of detector acceptance.  With this
cut the contribution of the non-resonant diagrams is again very small,
about 1\% over a large range in $\sqrt{s}$.  The difference between the
different schemes to include these diagrams is negligible, even near
threshold.  In this calculation we work in the 't Hooft-Feynman gauge.

\begin{figure}
\begin{picture}(554,504)(50,0)
\put(0,0){\strut\epsffile{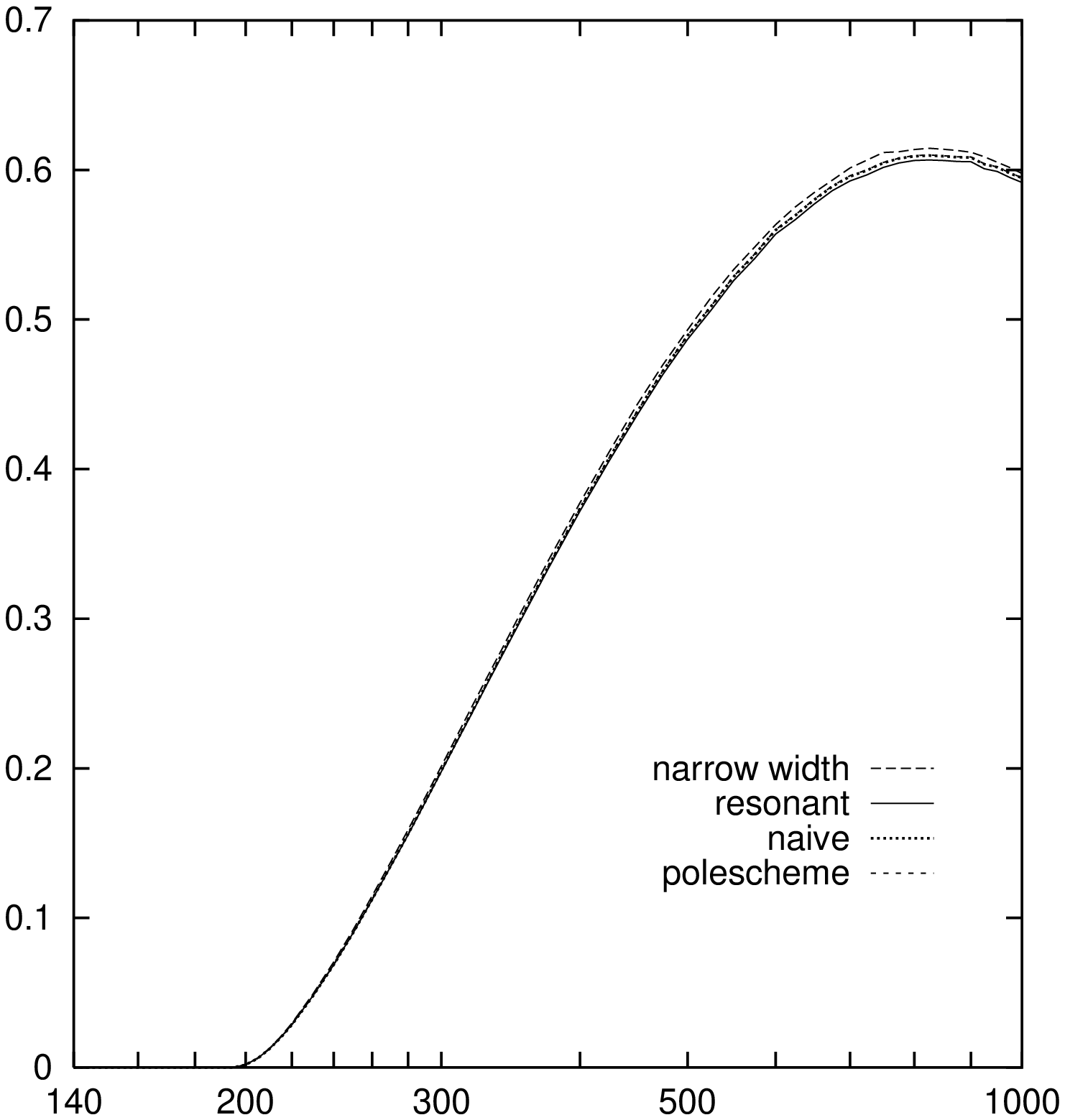}}
\put( 60.8,468.9){\makebox(0,0)[tr]{\shortstack{\Large\strut
$\sigma$\\\Large\strut$[\PB]$}}}
\put(480.8, 19.1){\makebox(0,0)[tr]{\Large$\sqrt{s_{ee}}[\GEV]$}}
\end{picture}
\caption[]{Cross sections for $\gamma\gamma\to \ell^- \bar{\nu}_\ell W^+$ in
the different schemes defined in the text with a more realistic photon
spectrum \cite{Ginsburg};
again we demand that $p_\perp(\ell) > 0.02\sqrt{s_{ee}}$}
\label{fig:polyWWabs}
\end{figure}

A more realistic picture is painted in Fig.\ \ref{fig:polyWWabs},
where we have folded the cross section with the lowest order spectrum
resulting from the conversion of the electron beam in a photon beam via
Compton scattering \cite{Ginsburg}.  The same conclusions hold here.


\section{Conclusions}

A study has been made of various schemes to handle processes with unstable
particles in intermediate states.  The application of these schemes in two
processes, $e^+e^- \to 4\mbox{ fermions}$ and $\gamma\gamma\to
e^-\bar{\nu}_eW^+$ shows that in general the corrections due to the width of
the $W$ boson are small, at least when the phase space is taken off-shell.

As was shown before,
for $e^+e^-$ collisions
the non-resonant diagrams become important only at
threshold (where the signal is low) and at high energies
($\sqrt{s}>500\GeV$).
At least in the commonly used gauges we worked in,
there is not much difference (less than 0.5\%) between
the various gauge invariant and gauge variant
schemes to include the non-resonant graphs.
This violation of $U(1)_{em}$ gauge invariance is
apparently not amplified by the large unitarity cancellations which occur in
these processes.  Near threshold we encounter some uncertainties.

For $\gamma\gamma$ collisions
the non-resonant graphs contribute much less, after a mild $p_\perp$ cut.
The difference does not amount to much more than 1\%\
over a large range of energies and even close to threshold.

\paragraph{Acknowledgements}

We would like to thank Daniel Wyler and Reinhold R\"uckl for useful
discussions.

\appendix


\clearpage
\section{Table}

\begin{table}[h]
{\small
\begin{tabular}{|l|l|l|l|l|}
\hline
name & kinematics & matrix element & gauge     & threshold behaviour \\
     &            &                & invariant & \\
\hline
narrow width & on-shell & $\displaystyle
R(m^2)$ & yes & undefined below threshold \\
resonant & off-shell & $\displaystyle
\frac{R(m^2)}{p^2-m^2+im\Gamma}$ & yes & zero
below threshold \\
offshell & off-shell & $\displaystyle
\frac{R(p^2)}{p^2-m^2+im\Gamma}$ & no  & wrong
below threshold \\
naive    & off-shell & $\displaystyle
\frac{R(p^2)}{p^2-m^2+im\Gamma} + N(p^2)$ & no  & ok \\
overall  & off-shell & $\displaystyle
\frac{R(p^2)}{p^2-m^2+im\Gamma} +
	\frac{p^2-m^2}{p^2-m^2+im\Gamma}N(p^2)$ & yes & ok \\
polescheme & off-shell & $\displaystyle
\frac{R(m^2)}{p^2-m^2+im\Gamma} +
	\tilde{R}(p^2) + N(p^2)$ & yes & wrong around threshold \\
\hline
\end{tabular}
}
\caption{A summary of the six different schemes to treat unstable
particles defined in section~\protect\ref{sec:method}}.
\label{table}
\end{table}


\section{Figures}

\begin{figure}
\begin{picture}(554,504)(50,0)
\put(0,0){\strut\epsffile{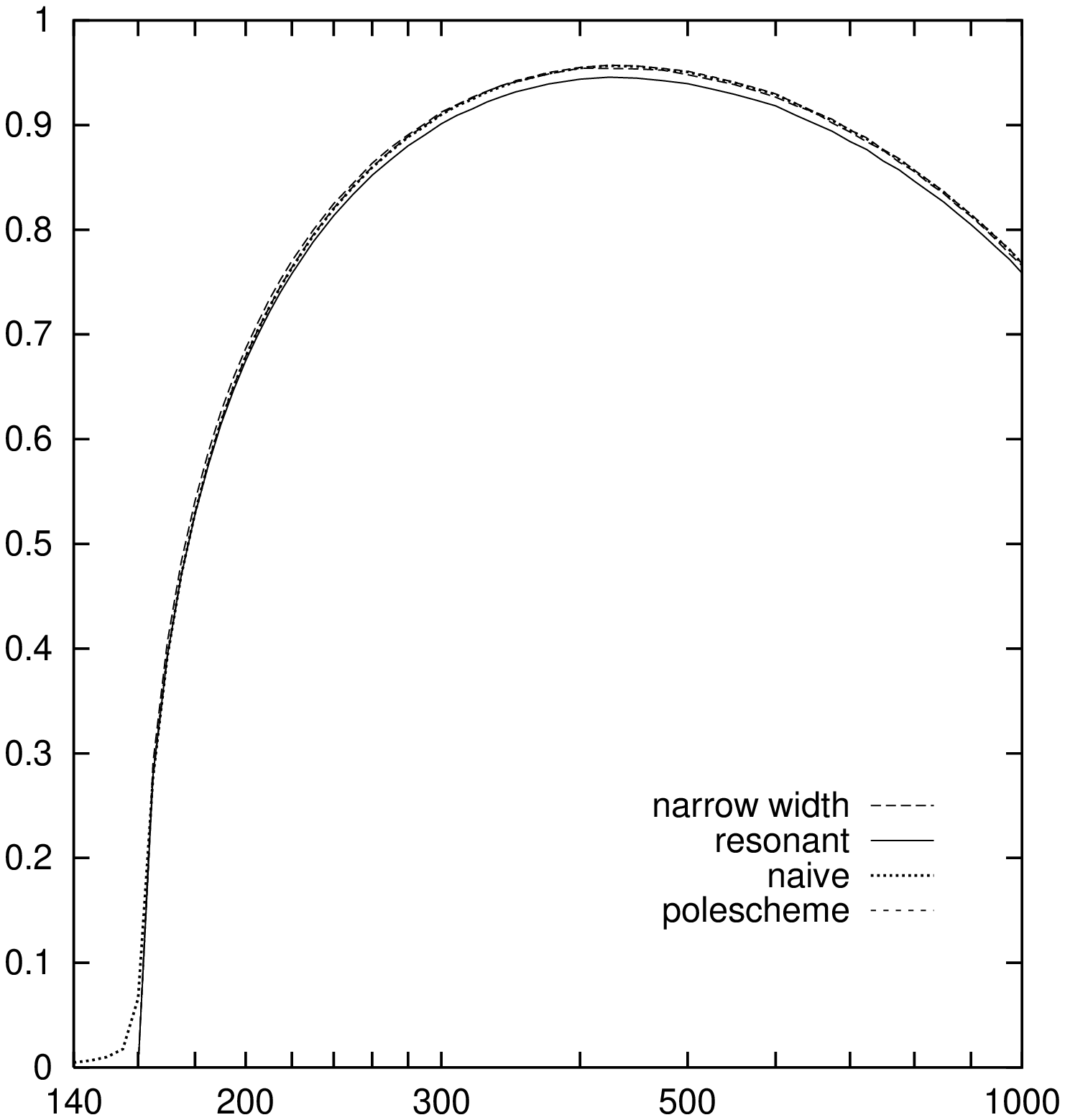}}
\put( 60.8,468.9){\makebox(0,0)[tr]{\shortstack{\Large\strut
$\sigma$\\\Large\strut$[\PB]$}}}
\put(480.8, 19.1){\makebox(0,0)[tr]{\Large$\sqrt{s_{\gamma\gamma}}[\GEV]$}}
\end{picture}
\caption[]{Cross sections for $\gamma\gamma\to \ell^- \bar{\nu}_\ell W^+$ in
the different schemes defined in the text, with $p_\perp(\ell) >
0.02\sqrt{s_{\gamma\gamma}}$ and a monochromatic laser beam.  The difference
between the ``naive'' and ``overall'' schemes is again not
visible on this plot.}
\label{fig:monoWWabs}
\end{figure}

\begin{figure}
\begin{picture}(554,504)(50,0)
\put(0,0){\strut\epsffile{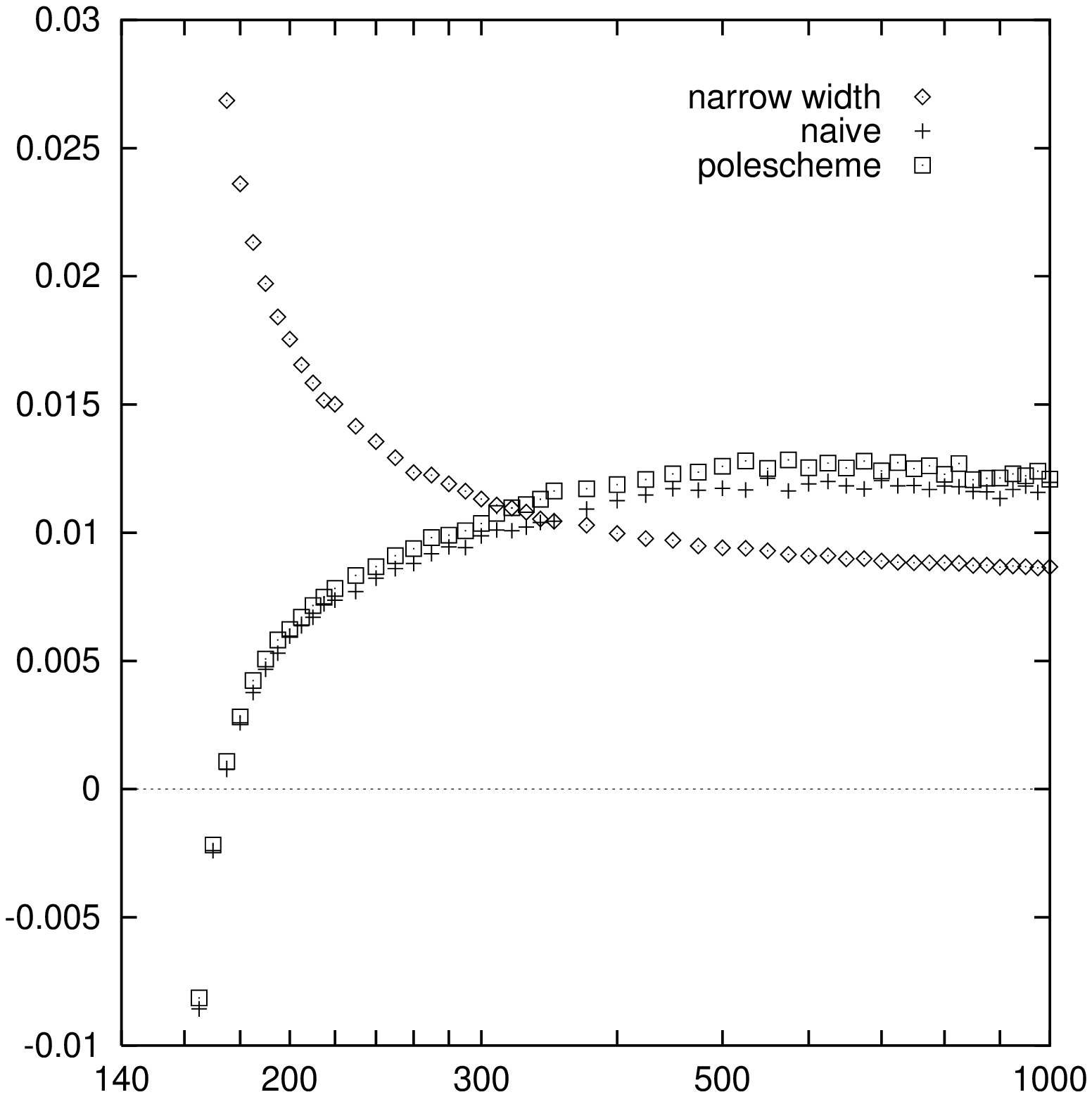}}
\put( 60.8,468.9){\makebox(0,0)[tr]{\Large\strut
$\frac{\displaystyle\sigma-\sigma_{\rm res}}{\displaystyle\sigma_{\rm res}}$}}
\put(480.8, 19.1){\makebox(0,0)[tr]{\Large$\sqrt{s_{\gamma\gamma}}[\GEV]$}}
\end{picture}
\caption[]{Differences with the doubly resonant cross section for
$\gamma\gamma\to \ell^- \bar{\nu}_\ell W^+$ as a fraction of the resonant
cross section.  The full $\order(\Gamma)$ corrections, including the width of
the $W^+$, are about twice this size.}
\label{fig:monoWWfrac}
\end{figure}

\begin{figure}
\begin{picture}(554,504)(50,0)
\put(0,0){\strut\epsffile{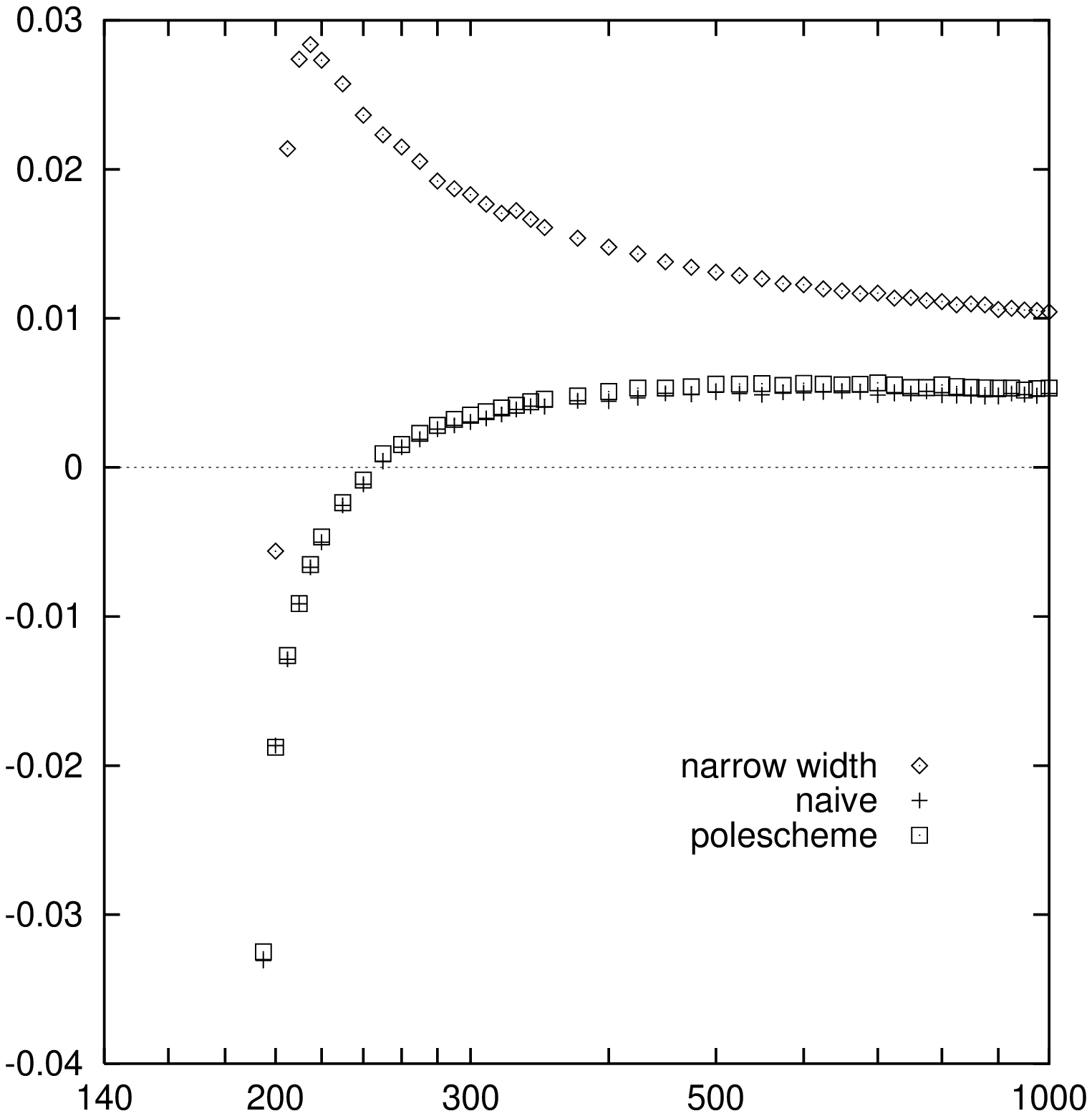}}
\put( 60.8,468.9){\makebox(0,0)[tr]{\Large$\frac{\displaystyle
\sigma-\sigma_{\rm res}}{\displaystyle\sigma_{\rm res}}$}}
\put(480.8, 19.1){\makebox(0,0)[tr]{\Large$\sqrt{s_{ee}}[\GEV]$}}
\end{picture}
\caption[]{Relative differences with the doubly resonant cross section for
$\gamma\gamma\to \ell^- \bar{\nu}_\ell W^+$
with a realistic photon spectrum \cite{Ginsburg}.
Again, the full $\order(\Gamma)$
corrections, including the width of the $W^+$, are about twice this size.}
\label{fig:polyWWfrac}
\end{figure}

\end{document}